# Perspectives for Spintronics in 2D Materials

Wei Han[1,2,3§]

[1]International Center for Quantum Materials, School of Physics, Peking University, Beijing 100871, P. R. China

[2]Collaborative Innovation Center of Quantum Matter, Beijing 100871, P. R. China

[3]Key Laboratory of Polar Materials and Devices, Ministry of Education, East China Normal University, Shanghai 200062, P. R. China

[§]Email: weihan@pku.edu.cn

**Abstract:**

The past decade has been especially creative for spintronics since the (re)discovery of various two dimensional (2D) materials. Due to the unusual physical characteristics, 2D materials have provided new platforms to probe the spin interaction with other degrees of freedom for electrons, as well as to be used for novel spintronics applications. This review briefly presents the most important recent and ongoing research for spintronics in 2D materials.

**Introduction**

Two dimensional (2D) materials are generally categorized as 2D allotropes of various elements or compounds, in which the electron transport is confined to a plane. Intrigued from the (re)discovery of graphene, isolating single atomic layers of van der Waals materials has been one of the most emerging research fields [1-4]. Different from their bulk materials, 2D materials have been shown to exhibit many festinating physical properties. Just name a few for example, room temperature quantum Hall effect has been observed in single layer graphene due to its unusual liner dispersion at the Dirac point [5,6]. The high electron/hole mobility in graphene and the metallic surface states of topological insulators (TIs) are very promising for future electronics devices [7-10]. And the unique physical properties in the 2D transition metal dichalcogenides (TMDCs) could be used for future valleytronics devices [11-17].

For spintronics, which aims to utilize the spin degree freedom of electrons for novel information storage and logic devices [18-21], these 2D materials are very attractive. The ultra-low spin orbit coupling in graphene already made it one of the most promising candidates for spin channel [22-29]. The unusual spin-momentum locking properties of the surface states in TIs provide a method to control the spin polarization via the charge current direction [10,29-36]. The unique spin-valley coupling in 2D TMDCs provides a platform to use valley for manipulating the spins [11,16]. The inueced magnetism into graphene or the surface states of TIs is particularly interesting towards the quantum anomalous Hall effect (QAHE) [27,29,37-50]. The enhanced spin orbit coupling in hydrogen doped graphene, silicene, germanane, tin and 2D TMDCs are potentially candidates for quantum spin Hall effect [27,51-54]. And the spin orbit torque at the TIs/ferromagnet interface has been demonstrated to be significantly larger than conventional heavy metals [55,56].

In this article, we briefly review the most recent developments of spintronics in 2D materials and this paper is organized as follows. The first section presents the current status of spin transport in 2D materials. The second section reviews of the induced magnetic moments in graphene and TIs by doping and proximity effect. The third section describes the spin/valley Hall effect and spin orbit torque, and the final section discusses the outlook of the research field of spintronics in 2D materials.

**Spin transport in 2D materials**

In 1990, Datta and Das proposed the idea of spin field effect transistor (spin FET), which relies on manipulating of the spins during transport in a semiconductor with an electric field [57]. Since then, enormous effort has been spent in injecting, detecting and manipulating spin polarized electrons in metals and conventional semiconductors [58-63]. However, in spite of these effort, an ideal candidate material for the spin channel is still needed. 2D materials, including graphene, the surface states of TIs, and TMDCs could be promising candidates due to their unusual spin dependent physical properties.

Graphene, a single or few layer of carbon atoms, exhibits very low spin-orbit coupling, ultra-high mobility and the gate tunable conductivities [1,2,64]. To electrically inject and detect spin polarized electrons in graphene, the nonlocal measurement geometry is mostly used. As shown in Fig. 1a, the spins are injected at the ferromagnetic (FM) electrode E2, and detected at the electrode E3. One typical nonlocal magnetoresistance (MR) curve is shown in Fig. 1c. The resistance difference between the parallel and anti-parallel states of the magnetization directions of E2 and E3 is called the nonlocal MR ($\Delta R_{NL}$). During the diffusion, a magnetic field perpendicular to the graphene device could precess the spins, and as a result of which, the detected spin voltage has a strong dependence on the applied magnetic field (Fig. 1b). This Hanle spin precession provides a way to determine of the spin lifetime and spin diffusion constant. As shown in Fig. 1d, the spin lifetime ($\tau_s$) and spin diffusion constant ($D$) are determined to be 771 ps and 0.020 m²/s, by numerical fitting based on the following equation [65]:

$$R_{NL} \propto \pm \int_0^\infty \frac{1}{\sqrt{4\pi Dt}} \exp\left[-\frac{L^2}{4Dt}\right] \cos(\omega_L t) \exp(-t/\tau_s) dt \quad (1)$$

where the + (-) sign is for the parallel (antiparallel) magnetization state, $L$ is the distance from injector (E2) to detector (E3), and $\varpi_L = g\mu_B H_\perp / \hbar$ is the Larmor frequency, in which $g$ is the g-factor, $\mu_B$ is the Bohr magneton, and $\hbar$ is the reduced Planck's constant. Details on the spin transport and spin precession in graphene have been discussed in several previous review articles [27,28,66]. Very intriguingly, long spin diffusion length of 24 micro meters has been observed in boron nitride encapsulated bilayer graphene recently [67].

For the metallic surface states of TIs, the spin lies in plane and is locked at the right angles of the momentum (Fig. 2a), which could be used to control the spin current via the charge current direction. Recently, Li et al used the tunneling FM contacts to probe the spin current while applying a charge current on the $Bi_2Se_3$ surface [33]. As shown in Fig. 2b, the carriers' momentum direction ($\vec{k}$) locks the spin direction to be in the $-\vec{y}$ direction, and this spin direction could be measured via the FM contacts (Fe). Fig. 2c and 2d show the magnetic field dependent voltage between the FM contacts and the Au contact for positive and negative 2 mA currents respectively. As the magnetization of the Fe electrode switches, the measured voltage shows a sharp increase or decrease depending on whether the Fe electrode magnetization direction and carrier spins' direction are parallel or anti-parallel to each other. Besides this $Bi_2Se_3$, several other TIs, including $(Bi_{0.53}Sb_{0.47})_2Te_3$, $Bi_{1.5}Sb_{0.5}Te_{1.7}Se_{1.3}$, $Bi_2Te_2Se$ have also exhibited this spin momentum locking property in their metallic surface states [34-36]. However, to ambitiously prove that this voltage is from the spin polarized surface states, further studies are needed. For example, the spin Hall effect in bulk states of TIs can also contribute to the spin current generation on the surface. Besides, the fringe-field-induced Hall voltages should be also examined in future studies [68].

The spin directions of the 2D TMDCs are different for the two valleys, K and K' [16]. Despite a lot of efforts, electrically injection and detection of the spin polarized current in 2D TMDCs have not been demonstrated yet. One of the main challenges is that the Schottky barrier resistance is very large, limiting the efficient spin injection [69,70]. A promising route to inject spin polarized carriers is to use thin MgO tunnel barrier between the $MoS_2$ and FM electrodes [71].

**Induced magnetism**

Induced magnetism in graphene and the surface states of TIs holds the potentials towards the realization of QAHE. Very intriguingly, the QAHE has been observed on magnetic TIs recently by several groups [44-46,72] at ultralow temperatures. Generally speaking, there are two routes to induce magnetism in a nonmagnetic material. The first one is doping another element, such as Mn, Cr et al, or creating defects. The most typical example is the diluted magnetic semiconductor, Mn doped GaAs [73,74]. The carriers exchange with local moment on Mn generates spin splitting in the band structure of GaAs, thus inducing magnetism. The other one is exchange coupling via proximity effect in adjacent with FM materials, with spin polarized *d* or *f* orbitals.

For graphene, hydrogen doping and vacancy defects have been used to induce magnetism. Paramagnetic moments were first observed in graphene with vacancy defects by SQUID measurement and later in graphene with hydrogen doping or vacancy defects probed by pure spin current [75,76]. In the later one, McCreary et al used hydrogen atomic source to dope the graphene (Fig. 3a) and performed in situ spin transport measurement. As shown in Fig. 3c, the dip at zero magnetic field for the nonlocal spin transport measurement indicates that the diffusive spins interact with the local hydrogen induced magnetic moments via exchanging coupling. Up to date, however, long range ferromagnetic order in doped graphene is still missing. On the other hand, proximity induced ferromagnetic graphene has been observed in the heterostructures of graphene and ferromagnetic insulator, yttrium iron garnet (YIG) [39]. In their study, the graphene device was transferred onto YIG thin films grown via pulsed laser deposition, and a top gate was used to tune the carrier density in graphene (Fig. 3b). As shown in Fig. 3d, anomalous Hall effect has been observed up to 250 K in graphene on YIG.

Quantum spin Hall effect has been observed in HgTe quantum well and inverted InAs/GaSb quantum wells [77,78]. By breaking the time reversal symmetry in TIs with magnetization, the conducting carriers with topological properties could take over the role of an external magnetic field, providing a route towards the QAHE (Fig. 4a) [40,41,44]. The first demonstration of this QAHE was achieved by Chang et al in Cr doped magnetic TI, $Cr_{0.15}(Bi_{0.1}Sb_{0.9})_{1.85}Te_3$. As shown in Fig. 4b and 4c, when the Fermi level is tuned into the TI bands, the anomalous Hall resistance shows a quantization of $(h/e^2)$ [44]. Another route towards the QAHE is by exchanging interaction with a magnetic insulator. Wei et al grew $Bi_2Se_3$ on EuS using molecular beam epitaxy, where EuS is a ferromagnetic insulator with the Curie temperature of ~ 16 K [47]. Anomalous Hall effect is observed due to the exchange interaction between the spin polarized 4$f$ orbitals of the EuS and the carriers in the $Bi_2Se_3$ conducting band (Fig. 4d and 4e).

**Spin/Valley Hall effect and Spin orbit torque**

Conventionally, the spin current is generated from the FM materials, in which the spin up and spin down polarized carriers are degenerated at the Fermi Level. A new mechanism to generate spin current is via spin Hall effect in a nonmagnetic material, which is first observed in a conducting GaAs channel [79,80].

The spin orbit coupling in graphene could be enhanced by doping [81]. Recently, A Hall bar device on hydrogen doped graphene exhibited large colossal spin Hall effect [82]. As shown in Fig. 5a and 5b, the current flows from $I_S$ to $I_D$, and the voltage is measured on the other two bars in a nonlocal geometry. This giant nonlocal resistance indicates the largely enhanced spin orbit coupling in graphene. However, there are also reports stating that this nonlocal voltage could be associated with some unknown mechanisms that are not related to spin [83].

The valley Hall effect has been observed in $MoS_2$ and symmetry broken graphene [17,84]. As shown in Fig. 5c, the valley degeneracy in $MoS_2$ is broken, and the two valleys exhibit different spin dependent properties. This spin valley coupling could be probed in the optical method via circular polarized light. As shown in Fig. 5d, the Hall voltage depends on whether the K or K' valley carriers are excited.

One of the many useful applications of the spin Hall effect is to manipulate the magnetization of the adjacent FM layer via spin orbit torque. The spin-momentum locking in the surface metallic states of TIs provides a unique way to generate efficient spin current. The first breakthrough is observation of very large spin orbit torque at the $Bi_2Se_3$-Py interface, in which the $Bi_2Se_3$ shows metallic behavior. Using the spin torque ferromagnetic resonance measurement (Fig. 6a), two distinct torques, namely, spin orbit torque and Oersted field torque are probed around the resonance condition of Py thin films [55]. As shown in Fig. 6b, the measured voltages have symmetric and antisymmetric components, of which the symmetric component is a result of the spin orbit torque and the antisymmetric component is due to the Orested field torque. The spin generation efficiency could be estimated from the ratio between these two torques. The effective spin Hall angle of $Bi_2Se_3$ is estimated to be 2.0 - 3.5, which is significant larger than the values reported in heavy metals, such as Pt, β-Ta, and β-W [55,85-88]. A following major breakthrough is the epitaxial integration of the topological insulator $(Bi_{0.5}Sb_{0.5})_2Te_3$, and the magnetic topological insulator, $(Cr_{0.08}Bi_{0.54}Sb_{0.38})_2Te_3$. As shown in Fig. 6c, anomalous Hall resistance is measured to probe the magnetization while a DC current is applied. When the magnetization flips its direction, the sign of the anomalous Hall effect switches, as shown in Fig. 6d. From there, an effective spin Hall angle is estimated to be almost three orders of magnitude larger than heavy metals [56].

**Summary and Outlook**

In summary, the spintronics in 2D materials is an exciting scientific research direction and has many potential applications for future technologies. Looking forward, there are many more opportunities.

1) Spin transport in novel 2D materails, including the TMDCs, silicene and germanane, phosephosre, and their heterosturatures [3,27].
2) High temperature robust quanmtum spin Hall and QAHE states. These include the optimal materials growth of the magnetically doped TIs [72], and the atomical layer eniginnering of the ferromagnetic materials to exchange coupled to the TIs [39,89]. Other material candiates including TMDCs [54], Tin based thin films [53,90,91], oxide interface [92,93] and Heuslter materials [94] could also be very attractive.

3) Towards larger spin orbital efficiency in quantum materials via enhanced spin orbit coupling in 2D materials and the novel mechanisms, such as valley current [16,95], to generate spin current beyond the spin Hall effect and Rashba spin orbit coupling [96]

4) The coupling of more than two types degrees of freedoms. For example, using electrical field or even ferroelectricity to tune the spin-valley coupling in TMDCs and oxide heterostructures [97,98].

**Acknowledgements**

W. H. acknowledges the help from Yangyang Chen and Wenyu Xing. W. H. also acknowledges the funding support from the National Basic Research Programs of China (2015CB921104 and 2014CB920902), the National Natural Science Foundation of China (NSFC Grant 11574006), and the 1000 Talents Program for Young Scientists of China.

**References:**

[1]     A. K. Geim and K. S. Novoselov, Nat. Mater. **6**, 183 (2007).

[2]     S. Das Sarma, S. Adam, E. H. Hwang, and E. Rossi, Rev. Mod. Phys. **83**, 407 (2011).

[3]     A. K. Geim and I. V. Grigorieva, Nature **499**, 419 (2013).

[4]     S. Z. Butler, S. M. Hollen, L. Cao, Y. Cui, J. A. Gupta, H. R. Gutiérrez, T. F. Heinz, S. S. Hong, J. Huang, A. F. Ismach, E. Johnston-Halperin, M. Kuno, V. V. Plashnitsa, R. D. Robinson, R. S. Ruoff, S. Salahuddin, J. Shan, L. Shi, M. G. Spencer, M. Terrones, W. Windl, and J. E. Goldberger, ACS Nano **7**, 2898 (2013).

[5]     K. S. Novoselov, A. K. Geim, S. V. Morozov, D. Jiang, M. I. Katcnelson, I. V. Grigorieva, S. V. Dubonos, and A. A. Firsov, Nature **438**, 197 (2005).

[6]     Y. Zhang, Y.-W. Tan, H. L. Stormer, and P. Kim, Nature **438**, 201 (2005).

[7] S. V. Morozov, K.S.Novoselov, M. I. Katsnelson, F. Schedin, D. C. Elias, J. A. Jaszczak, and A. K. Geim, Phys. Rev. Lett. **100**, 016602 (2008).

[8] D.-X. Qu, Y. S. Hor, J. Xiong, R. J. Cava, and N. P. Ong, Science **329**, 821 (2010).

[9] D. Kim, S. Cho, N. P. Butch, P. Syers, K. Kirshenbaum, S. Adam, J. Paglione, and M. S. Fuhrer, Nat. Phys. **8**, 460 (2012).

[10] M. Z. Hasan and C. L. Kane, Rev. Mod. Phys. **82**, 3045 (2010).

[11] D. Xiao, G.-B. Liu, W. Feng, X. Xu, and W. Yao, Phys. Rev. Lett. **108**, 196802 (2012).

[12] H. Zeng, J. Dai, W. Yao, D. Xiao, and X. Cui, Nat. Nanotechnol. **7**, 490 (2012).

[13] K. F. Mak, K. He, J. Shan, and T. F. Heinz, Nat. Nanotechnol. **7**, 494 (2012).

[14] T. Cao, G. Wang, W. Han, H. Ye, C. Zhu, J. Shi, Q. Niu, P. Tan, E. Wang, B. Liu, and J. Feng, Nat. Commun. **3**, 887 (2012).

[15] S. Wu, J. S. Ross, G.-B. Liu, G. Aivazian, A. Jones, Z. Fei, W. Zhu, D. Xiao, W. Yao, D. Cobden, and X. Xu, Nat. Phys. **9**, 149 (2013).

[16] X. Xu, W. Yao, D. Xiao, and T. F. Heinz, Nat. Phys. **10**, 343 (2014).

[17] K. F. Mak, K. L. McGill, J. Park, and P. L. McEuen, Science **344**, 1489 (2014).

[18] S. A. Wolf, D. D. Awschalom, R. A. Buhrman, J. M. Daughton, S. von Molnar, M. L. Roukes, A. Y. Chtchelkanova, and D. M. Treger, Science **294**, 1488 (2001).

[19] I. Zutic, J. Fabian, and S. Das Sarma, Rev. Mod. Phys **76**, 323 (2004).

[20] A. Fert, Rev. Mod. Phys. **80**, 1517 (2008).

[21] S. D. Bader and S. S. P. Parkin, Annu. Rev. Cond. Mat. **1**, 71 (2010).

[22] N. Tombros, C. Jozsa, M. Popinciuc, H. T. Jonkman, and B. J. van Wees, Nature **448**, 571 (2007).

[23] W. Han, K. Pi, K. M. McCreary, Y. Li, J. J. I. Wong, A. G. Swartz, and R. K. Kawakami, Phys. Rev. Lett. **105**, 167202 (2010).

[24] W. Han and R. K. Kawakami, Phys. Rev. Lett. **107**, 047207 (2011).

[25] B. Dlubak, M.-B. Martin, C. Deranlot, B. Servet, S. Xavier, R. Mattana, M. Sprinkle, C. Berger, W. A. De Heer, F. Petroff, A. Anane, P. Seneor, and A. Fert, Nat. Phys. **8**, 557 (2012).

[26] M. H. D. Guimarães, P. J. Zomer, J. Ingla-Aynés, J. C. Brant, N. Tombros, and B. J. v. Wees, Phys. Rev. Lett. **113**, 086602 (2014).

[27]   W. Han, R. K. Kawakami, M. Gmitra, and J. Fabian, Nat. Nanotechnol. **9**, 794 (2014).

[28]   R. Stephan, Å. Johan, B. Bernd, C. Jean-Christophe, C. Mairbek, D. Saroj Prasad, D. Bruno, F. Jaroslav, F. Albert, G. Marcos, G. Francisco, G. Irina, S. Christian, S. Pierre, S. Christoph, O. V. Sergio, W. Xavier, and W. Bart van, 2D Materials **2**, 030202 (2015).

[29]   D. Pesin and A. H. MacDonald, Nat. Mater. **11**, 409 (2012).

[30]   X.-L. Qi and S.-C. Zhang, Rev. Mod. Phys. **83**, 1057 (2011).

[31]   D. Hsieh, Y. Xia, D. Qian, L. Wray, J. H. Dil, F. Meier, J. Osterwalder, L. Patthey, J. G. Checkelsky, N. P. Ong, A. V. Fedorov, H. Lin, A. Bansil, D. Grauer, Y. S. Hor, R. J. Cava, and M. Z. Hasan, Nature **460**, 1101 (2009).

[32]   P. Roushan, J. Seo, C. V. Parker, Y. S. Hor, D. Hsieh, D. Qian, A. Richardella, M. Z. Hasan, R. J. Cava, and A. Yazdani, Nature **460**, 1106 (2009).

[33]   C. H. Li, O. M. J. van't Erve, J. T. Robinson, Y. Liu, L. Li, and B. T. Jonker, Nat. Nanotechnol. **9**, 218 (2014).

[34]   J. Tang, L.-T. Chang, X. Kou, K. Murata, E. S. Choi, M. Lang, Y. Fan, Y. Jiang, M. Montazeri, W. Jiang, Y. Wang, L. He, and K. L. Wang, Nano Letters **14**, 5423 (2014).

[35]   Y. Ando, T. Hamasaki, T. Kurokawa, K. Ichiba, F. Yang, M. Novak, S. Sasaki, K. Segawa, Y. Ando, and M. Shiraishi, Nano Letters **14**, 6226 (2014).

[36]   J. Tian, I. Miotkowski, S. Hong, and Y. P. Chen, Sci. Rep. **5**, 14293 (2015).

[37]   W.-K. Tse, Z. Qiao, Y. Yao, A. H. MacDonald, and Q. Niu, Phys. Rev. B **83**, 155447 (2011).

[38]   H. Zhang, C. Lazo, S. Blügel, S. Heinze, and Y. Mokrousov, Phys. Rev. Lett. **108**, 056802 (2012).

[39]   Z. Wang, C. Tang, R. Sachs, Y. Barlas, and J. Shi, Phys. Rev. Lett. **114**, 016603 (2015).

[40]   C.-X. Liu, X.-L. Qi, X. Dai, Z. Fang, and S.-C. Zhang, Phys. Rev. Lett. **101**, 146802 (2008).

[41]   R. Yu, W. Zhang, H.-J. Zhang, S.-C. Zhang, X. Dai, and Z. Fang, Science **329**, 61 (2010).

[42]   D. Culcer and S. Das Sarma, Phys. Rev. B **83**, 245441 (2011).

[43]   J. Wang and D. Culcer, Phys. Rev. B **88**, 125140 (2013).

[44]   C.-Z. Chang, J. Zhang, X. Feng, J. Shen, Z. Zhang, M. Guo, K. Li, Y. Ou, P. Wei, L.-L. Wang, Z.-Q. Ji, Y. Feng, S. Ji, X. Chen, J. Jia, X. Dai, Z. Fang, S.-C. Zhang, K. He, Y. Wang, L. Lu, X.-C. Ma, and Q.-K. Xue, Science **340**, 167 (2013).

[45]   X. Kou, S.-T. Guo, Y. Fan, L. Pan, M. Lang, Y. Jiang, Q. Shao, T. Nie, K. Murata, J. Tang, Y. Wang, L. He, T.-K. Lee, W.-L. Lee, and K. L. Wang, Phys. Rev. Lett. **113**, 137201 (2014).

[46]   J. G. Checkelsky, R. Yoshimi, A. Tsukazaki, K. S. Takahashi, Y. Kozuka, J. Falson, M. Kawasaki, and Y. Tokura, Nat. Phys. **10**, 731 (2014).

[47]   P. Wei, F. Katmis, B. A. Assaf, H. Steinberg, P. Jarillo-Herrero, D. Heiman, and J. S. Moodera, Phys. Rev. Lett. **110**, 186807 (2013).

[48]   L. D. Alegria, H. Ji, N. Yao, J. J. Clarke, R. J. Cava, and J. R. Petta, Appl. Phys. Lett. **105**, 053512 (2014).

[49]   M. Lang, M. Montazeri, M. C. Onbasli, X. Kou, Y. Fan, P. Upadhyaya, K. Yao, F. Liu, Y. Jiang, W. Jiang, K. L. Wong, G. Yu, J. Tang, T. Nie, L. He, R. N. Schwartz, Y. Wang, C. A. Ross, and K. L. Wang, Nano Letters **14**, 3459 (2014).

[50]   Z. Jiang, C.-Z. Chang, C. Tang, P. Wei, J. S. Moodera, and J. Shi, Nano Letters **15**, 5835 (2015).

[51]   C. L. Kane and E. J. Mele, Phys. Rev. Lett. **95**, 226801 (2005).

[52]   C.-C. Liu, W. Feng, and Y. Yao, Phys. Rev. Lett. **107**, 076802 (2011).

[53]   Y. Xu, B. Yan, H.-J. Zhang, J. Wang, G. Xu, P. Tang, W. Duan, and S.-C. Zhang, Phys. Rev. Lett. **111**, 136804 (2013).

[54]   X. Qian, J. Liu, L. Fu, and J. Li, Science **346**, 1344 (2014).

[55]   A. R. Mellnik, J. S. Lee, A. Richardella, J. L. Grab, P. J. Mintun, M. H. Fischer, A. Vaezi, A. Manchon, E. A. Kim, N. Samarth, and D. C. Ralph, Nature **511**, 449 (2014).

[56]   Y. Fan, P. Upadhyaya, X. Kou, M. Lang, S. Takei, Z. Wang, J. Tang, L. He, L.-T. Chang, M. Montazeri, G. Yu, W. Jiang, T. Nie, R. N. Schwartz, Y. Tserkovnyak, and K. L. Wang, Nat. Mater. **13**, 699 (2014).

[57]   S. Datta and B. Das, Appl. Phys. Lett. **56**, 665 (1990).

[58]   M. Johnson and R. H. Silsbee, Phys. Rev. Lett. **55**, 1790 (1985).

[59]   F. J. Jedema, A. T. Filip, and B. J. van Wees, Nature **410**, 345 (2001).

[60]   T. Kimura and Y. Otani, Phys. Rev. Lett. **99**, 196604 (2007).

[61]   B. T. Jonker, G. Kioseoglou, A. T. Hanbicki, C. H. Li, and P. E. Thompson, Nat. Phys. **3**, 542 (2007).

[62]   I. Appelbaum, B. Huang, and D. J. Monsma, Nature **447**, 295 (2007).

[63] X. Lou, C. Adelmann, S. A. Crooker, E. S. Garlid, J. Zhang, K. S. M. Reddy, S. D. Flexner, C. J. Palmstrom, and P. A. Crowell, Nat. Phys. **3**, 197 (2007).

[64] A. H. Castro Neto, F. Guinea, N. M. R. Peres, K. S. Novoselov, and A. K. Geim, Rev. Mod. Phys. **81**, 109 (2009).

[65] F. J. Jedema, H. B. Heersche, A. T. Filip, J. J. A. Baselmans, and B. J. van Wees, Nature **416**, 713 (2002).

[66] W. Han, K. M. McCreary, K. Pi, W. H. Wang, Y. Li, H. Wen, J. R. Chen, and R. K. Kawakami, J. Magn. Magn. Mater. **324**, 369 (2012).

[67] J. Ingla-Aynés, M. H. D. Guimarães, R. J. Meijerink, P. J. Zomer, and B. J. van Wees, Phys. Rev. B **92**, 201410 (2015).

[68] E. K. de Vries, A. M. Kamerbeek, N. Koirala, M. Brahlek, M. Salehi, S. Oh, B. J. van Wees, and T. Banerjee, Phys. Rev. B **92**, 201102 (2015).

[69] A. Fert and H. Jaffrès, Phys. Rev. B **64**, 184420 (2001).

[70] A. Allain, J. Kang, K. Banerjee, and A. Kis, Nat. Mater. **14**, 1195 (2015).

[71] J.-R. Chen, P. M. Odenthal, A. G. Swartz, G. C. Floyd, H. Wen, K. Y. Luo, and R. K. Kawakami, Nano Letters **13**, 3106 (2013).

[72] C.-Z. Chang, W. Zhao, D. Y. Kim, H. Zhang, B. A. Assaf, D. Heiman, S.-C. Zhang, C. Liu, M. H. W. Chan, and J. S. Moodera, Nat. Mater. **14**, 473 (2015).

[73] T. Dietl, Nat. Mater. **9**, 965 (2010).

[74] T. Dietl and H. Ohno, Rev. Mod. Phys. **86**, 187 (2014).

[75] R. R. Nair, M. Sepioni, I. L. Tsai, O. Lehtinen, J. Keinonen, A. V. Krasheninnikov, T. Thomson, A. K. Geim, and I. V. Grigorieva, Nat. Phys. **8**, 199 (2012).

[76] K. M. McCreary, A. G. Swartz, W. Han, J. Fabian, and R. K. Kawakami, Phys. Rev. Lett. **109**, 186604 (2012).

[77] M. König, S. Wiedmann, C. Brüne, A. Roth, H. Buhmann, L. W. Molenkamp, X.-L. Qi, and S.-C. Zhang, Science **318**, 766 (2007).

[78] I. Knez, R.-R. Du, and G. Sullivan, Phys. Rev. Lett. **107**, 136603 (2011).

[79] D. Awschalom and N. Samarth, Phsyics **2**, 50 (2009).

[80] Y. K. Kato, R. C. Myers, A. C. Gossard, and D. D. Awschalom, Science **306**, 1910 (2004).

[81] A. H. Castro Neto and F. Guinea, Phys. Rev. Lett. **103**, 026804 (2009).

[82] J. Balakrishnan, G. Kok Wai Koon, M. Jaiswal, A. H. Castro Neto, and B. Ozyilmaz, Nat. Phys. **9**, 284 (2013).

[83] A. A. Kaverzin and B. J. van Wees, Phys. Rev. B **91**, 165412 (2015).

[84] R. V. Gorbachev, J. C. W. Song, G. L. Yu, A. V. Kretinin, F. Withers, Y. Cao, A. Mishchenko, I. V. Grigorieva, K. S. Novoselov, L. S. Levitov, and A. K. Geim, Science **346**, 448 (2014).

[85] L. Liu, T. Moriyama, D. C. Ralph, and R. A. Buhrman, Phys. Rev. Lett. **106**, 036601 (2011).

[86] W. Zhang, W. Han, X. Jiang, S.-H. Yang, and S. S. P. Parkin, Nat. Phys. **11**, 496 (2015).

[87] L. Liu, C.-F. Pai, Y. Li, H. W. Tseng, D. C. Ralph, and R. A. Buhrman, Science **336**, 555 (2012).

[88] C.-F. Pai, L. Liu, Y. Li, H. W. Tseng, D. C. Ralph, and R. A. Buhrman, Appl. Phys. Lett. **101**, 122404 (2012).

[89] W. Yuan, Y. Zhao, C. Tang, T. Su, Q. Song, J. Shi, and W. Han, Appl. Phys. Lett. **107**, 022404 (2015).

[90] S.-C. Wu, G. Shan, and B. Yan, Phys. Rev. Lett. **113**, 256401 (2014).

[91] F.-f. Zhu, W.-j. Chen, Y. Xu, C.-l. Gao, D.-d. Guan, C.-h. Liu, D. Qian, S.-C. Zhang, and J.-f. Jia, Nat. Mater. **14**, 1020 (2015).

[92] D. Xiao, W. Zhu, Y. Ran, N. Nagaosa, and S. Okamoto, Nat. Commun. **2**, 596 (2011).

[93] H. Zhang, J. Wang, G. Xu, Y. Xu, and S.-C. Zhang, Phys. Rev. Lett. **112**, 096804 (2014).

[94] S. Chadov, X. Qi, J. Kübler, G. H. Fecher, C. Felser, and S. C. Zhang, Nat. Mater. **9**, 541 (2010).

[95] V. T. Renard, B. A. Piot, X. Waintal, G. Fleury, D. Cooper, Y. Niida, D. Tregurtha, A. Fujiwara, Y. Hirayama, and K. Takashina, Nat. Commun. **6** (2015).

[96] A. Manchon, H. C. Koo, J. Nitta, S. M. Frolov, and R. A. Duine, Nat. Mater. **14**, 871 (2015).

[97] H. Yuan, M. S. Bahramy, K. Morimoto, S. Wu, K. Nomura, B.-J. Yang, H. Shimotani, R. Suzuki, M. Toh, C. Kloc, X. Xu, R. Arita, N. Nagaosa, and Y. Iwasa, Nat. Phys. **9**, 563 (2013).

[98] K. Yamauchi, P. Barone, T. Shishidou, T. Oguchi, and S. Picozzi, Phys. Rev. Lett. **115**, 037602 (2015).

**Figure Captions**

**Figure 1. Spin transport and relaxation in graphene.** (a) Nonlocal spin transport. E2 and E3 are ferromagnetic electrodes. (b) Hanle effect of the spin precession by a perpendicular magnetic field. (c) Nonlocal magnetoresistance measured on a typical graphene nonlocal spin valve with tunneling contacts. (d) Hanle spin precession measurement for parallel and anti-parallel configurations. Reproduced with permission from Nat. Nanotechnol. 9, 794 (2014). Copyright 2014 Nature Publishing Group.

**Figure 2. Spin transport in the surface metallic states of TIs.** (a) Schematic drawing for spin momentum locking in the Dirac band of TI surface states. (b) Concept drawing for the TI spin transport device structure, where FM electrode is used to probe the spin polarization of the conducting carriers. (c-d) The spin dependent voltage as a function of the magnetic field for positive and negative 2 mA currents, respectively. Reproduced with permission from Nat. Nanotechnol. 9, 218 (2014). Copyright 2014 Nature Publishing Group.

**Figure 3. Induced magnetism in graphene.** (a) Schematic illustration for hydrogen doped graphene. (b) Concept drawing for graphene exchange coupled to an atomically flat yttrium iron garnet ferromagnetic thin film. (c) The detection of paramagnetic moments in hydrogen doped graphene via pure spin current. (d) The anomalous Hall resistance measurements on magnetic graphene at various temperatures. (a) and (c) reproduced with permission from Phys. Rev. Lett. 109, 186604 (2012). Copyright 2012 American Chemical Society. (b) and (d) reproduced with permission from Phys. Rev. Lett. **114**, 016603 (2015). Copyright 2015 American Chemical Society.

**Figure 4. Induced magnetism in TIs.** (a) Quantum anomalous Hall effect for magnetic TI films. (b-c) The quantized anomalous Hall effect measured in $Cr_{0.15}(Bi_{0.1}Sb_{0.9})_{1.85}Te_3$ films. (d-e) The observation of anomalous Hall effect in magnetic $Bi_2Se_3$ films via magnetic proximity coupling to the EuS, a ferromagnetic insulator. (a-c) reproduced with permission from Science 340, 167 (2013). Copyright 2013 The American Association for the Advancement of Science. (d- e) reproduced with permission from Phys. Rev. Lett. 110, 186807 (2013). Copyright 2013 American Chemical Society.

**Figure 5. Spin/Valley Hall effects in 2D materials**. (a-b) The colossal spin Hall effect observed on hydrogen doped graphene via the Hall bar geometry. (c-d) The valley hall effect observed on $MoS_2$. (a-b) reproduced with permission from Nat. Phys. 9, 284 (2013). Copyright 2013 Nature Publishing Group. (c-d) reproduced with permission from Science 344, 1489 (2014). Copyright 2014 The American Association for the Advancement of Science.

**Figure 6. Giant spin orbit torque arising from topological insulators.** (a-b) Spin torque ferromagnetic resonance measurement in $Bi_2Se_3$/Py bilayer structure. (c-d) Magnetization switching measurements on magnetic TI/TI interface. The red and blue lines indicate the positive and negative currents. (a-b) reproduced with permission from Nature **511**, 449 (2014). Copyright 2014 Nature Publishing Group. (c-d) reproduced with permission from Nat. Mater. 13, 699 (2014). Copyright 2014 Nature Publishing Group.

Figure 1

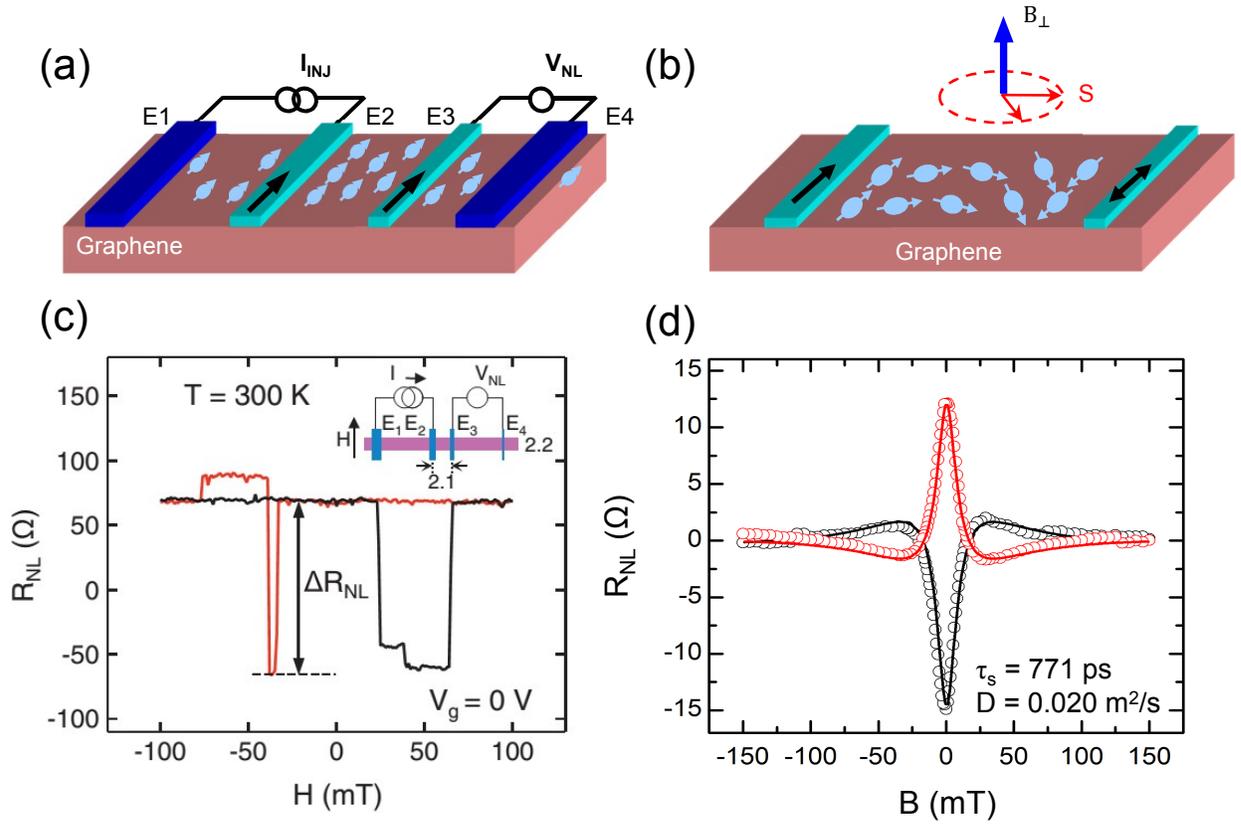

Figure 2

Figure 3

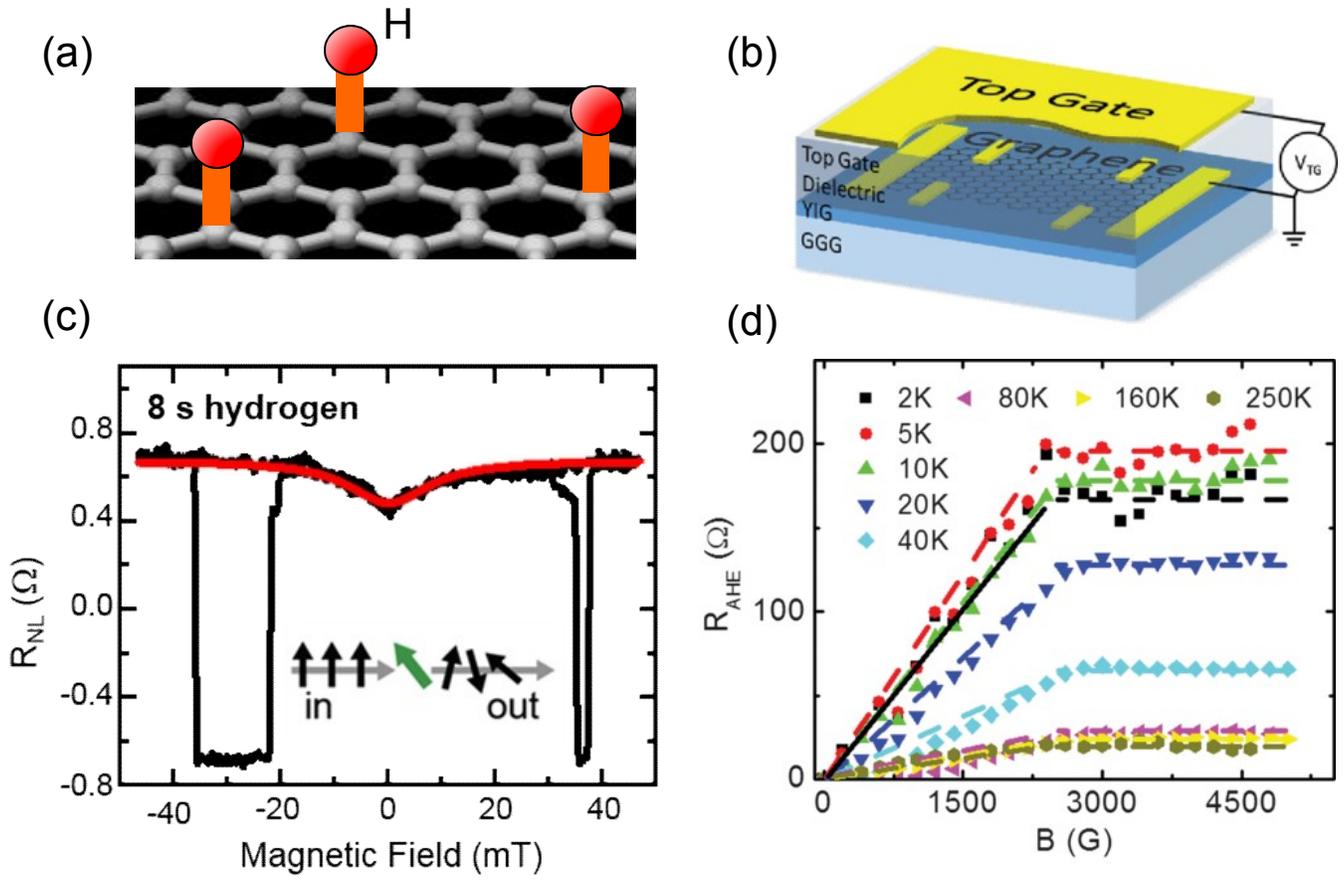

Figure 4

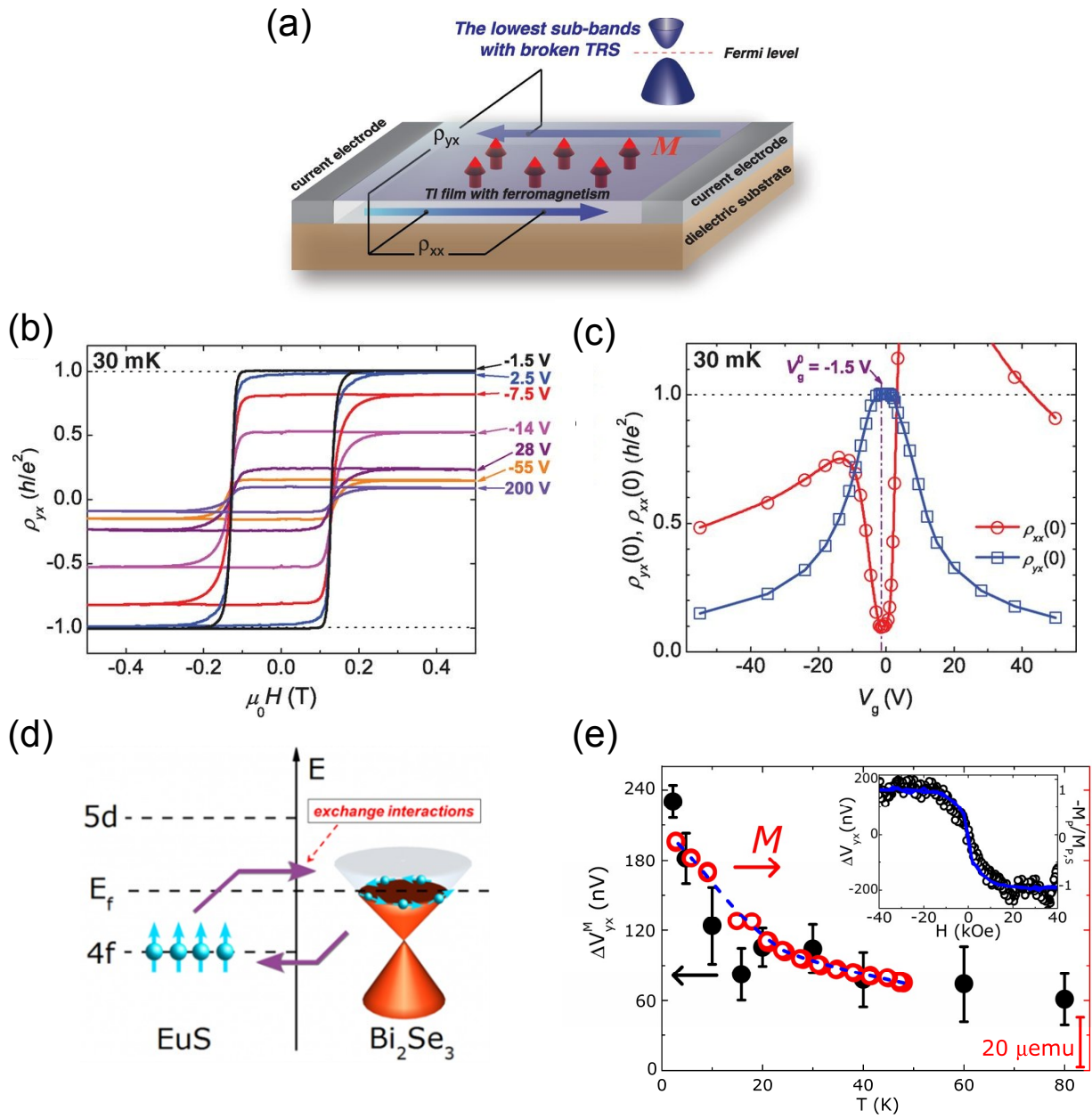

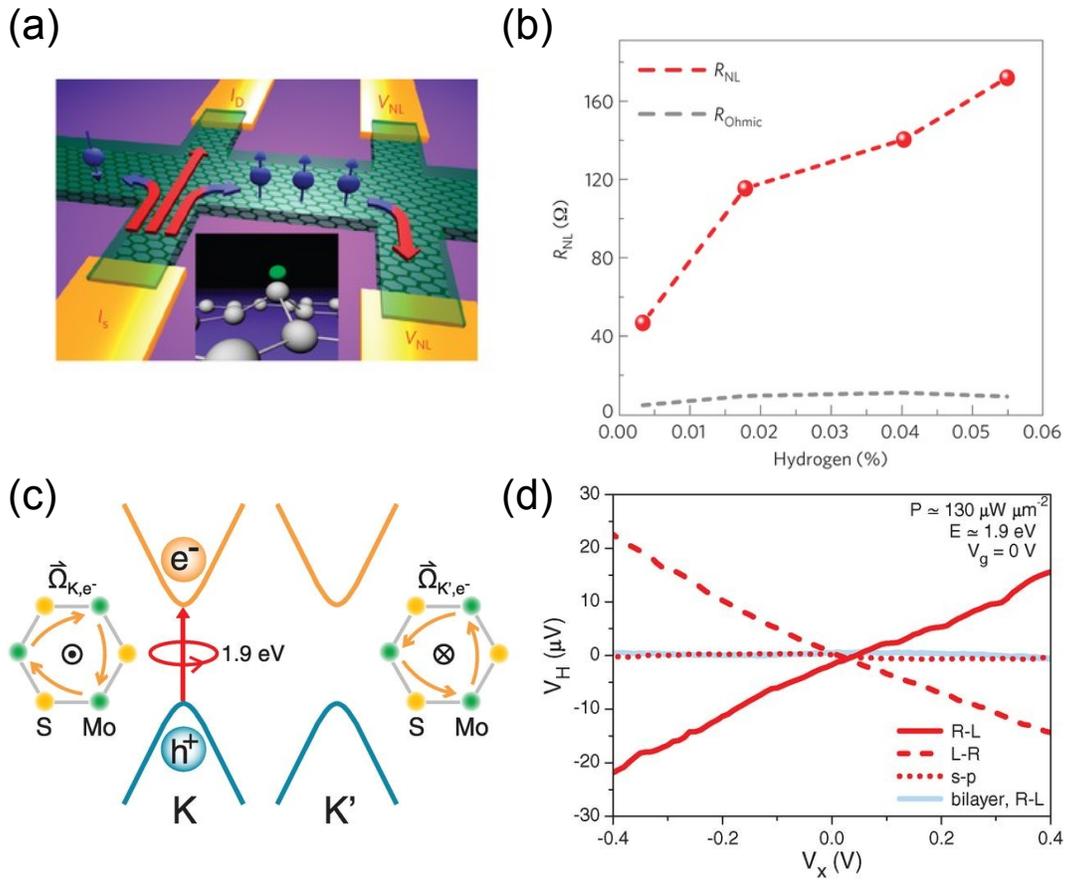

Figure 6

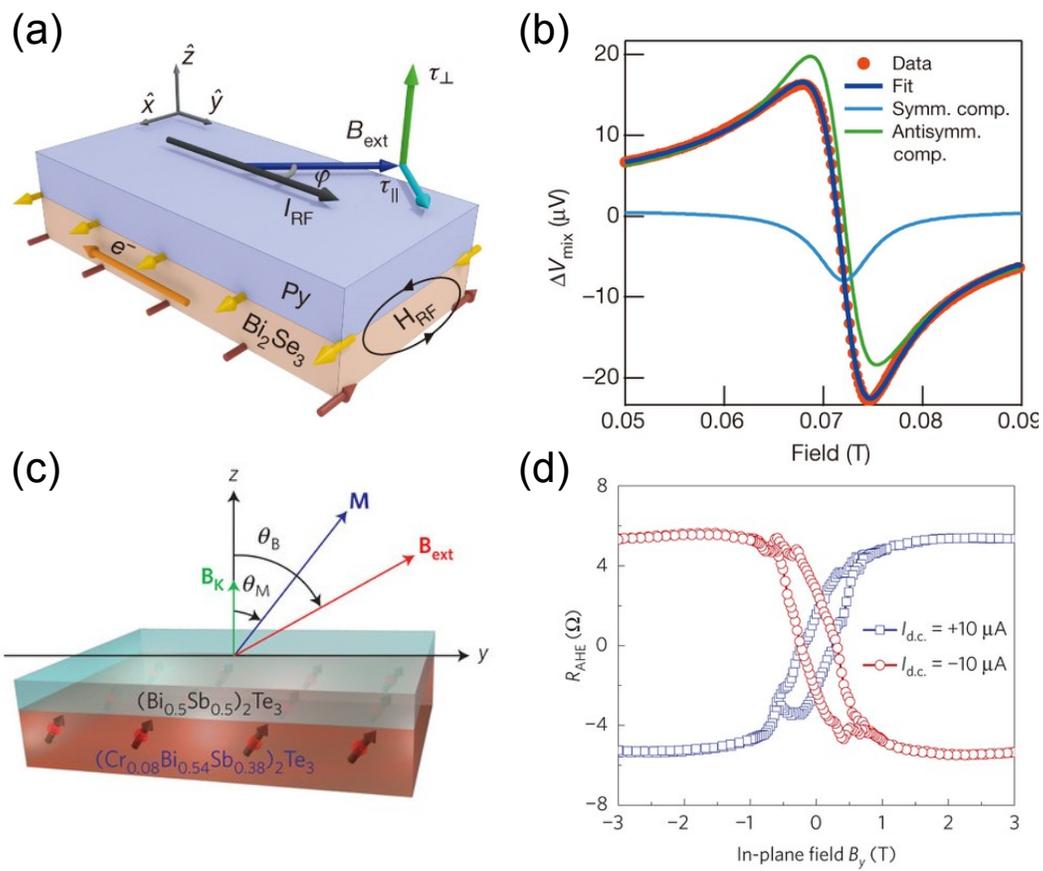